\documentclass[useAMS,usenatbib]{mnras}

\usepackage{graphicx}
\usepackage{amsmath}
\usepackage{amssymb}

\title[Determination of magnetic field strength on the event horizon of supermassive black holes...]{Determination of magnetic field strength on the event horizon of supermassive black holes in active galactic nuclei}

\author[M.Yu. Piotrovich et al.]{
M.Yu. Piotrovich$^1$ \thanks{E-mail: mpiotrovich@mail.ru},
A.G. Mikhailov$^2$ \thanks{E-mail: mag10629@yandex.ru},
S.D. Buliga$^1$
T.M. Natsvlishvili$^1$,
\\
$^1$ Central Astronomical Observatory at Pulkovo, 196140, Saint-Petersburg, Russia\\
$^2$ Special Astrophysical Observatory, 369167, Nizhnij Arkhyz, Zelenchukskiy region, Karachai-Cherkessian Republic, Russia}

\date{Accepted for publication in MNRAS}

\pubyear{2020}

\begin{document}

\label{firstpage}
\pagerange{\pageref{firstpage}--\pageref{lastpage}}
\maketitle

\begin{abstract}
We estimated the magnetic field strength at the event horizon for a sample of supermassive black holes (SMBHs) in active galactic nuclei (AGNs). Our estimates were made using the values of the inclination angles of the accretion disk to the line of sight, that we obtained previously from spectropolarimetric observations in the visible spectrum. We also used published values of full width at half maximum (FWHM) of spectral line $H_\beta$ from broad line region, masses of SMBHs and luminosity of AGNs at 5100 angstrom. In addition we used literature data on the spins of SMBHs obtained from their X-ray spectra. Our estimates showed that the magnetic field strength at the event horizon of the majority of SMBHs in AGNs ranges from several to tens of kG and have mean values of about $10^4$G. At the same time, for individual objects, the fields are significantly larger - of the order of hundreds kG or even 1 MG.
\end{abstract}


\section{Introduction} 

According to generally accepted notions, active galactic nuclei (AGNs) often have a magnetized accretion disk. It is believed that the magnetic field is formed as a result of the interaction of the accreting matter with the rotating SMBH \citep{moderski97, li02, wang02, wang03, zhang05, ma07}. Determining the radius dependence of the magnetic field in the accretion disk is a rather difficult task with many parameters (see, for example, the papers \citet{pariev03, gnedin12, baczko16}). These parameters include: the mass of the black hole $M_ {BH}$, the spin $a$ (dimensionless angular momentum of rotation $a = c J / G M_ {BH}^2$, where $J$ is the angular momentum of rotation of the black hole, $M_{BH}$ is its mass, $c$ is the speed of light, $G$ is the gravitational constant), the angle $i$ between the line of sight and the normal to the plane of the accretion disk and the radiation efficiency $\varepsilon$. Here, we estimated the magnetic field strength at the event horizon of a series of SMBHs using the angles $i$ obtained in our previous study \citep{afanasiev18}, as well as published data on X-ray spectra \citep{brenneman13, reynolds14}, on the value of $FWHM$ (full width at half amplitude) for the spectral emission line $H_\beta$ from the broad line region (BLR) and the luminosity values at 5100 angstroms $L_{5100}$ \citep{vestergaard06, feng14, sani10, sheinis17, decarli11, grier17, bentz13, kollatschny18, grupe15}.

\section{Calculation method} 

Recently there has been a discussion about problem of discrepancies of theoretically calculated sizes of classical thin accretion disk with the results of observations by echo mapping and microlensing methods. Analysis of the observational data shows that the size of the central region of accretion disk is several times bigger. This circumstance was an incentive for the development of models in which outflows/winds of the matter in the central region of accretion disk are considered. Such models explain both SED and large sizes of accretion disk by the selection of appropriate parameters \citep{laor14,slone12,edelson15,capellupo16,kokubo18,sun19}. However, there is the work \citet{homayouni19}, in which data analysis was carried out via echo mapping and they concluded that the best fit sizes of accretion disk are consistent with the Shakura-Sunyaev model within $1.5 \sigma$. Also, authors of \citet{yu20} concluded that accretion disk sizes are consistent with the theoretical predictions of the standard thin disk model, given the effects of disk variability.

In this paper, we consider the Shakura-Sunyaev model of the geometrically thin, optically thick disk \citep{shakura73}. For disks of this type, it was assumed in \citet{moderski97} that the magnetic pressure $P_ {magn} = B_H ^ 2/8 \pi$ and the pressure of the accreting matter $P_{acc} = c \dot{M} / 4 \pi R_H^2$ on the event horizon are equal. Here $B_H$ is the magnetic field on the event horizon, $\dot{M}$ the accretion rate, $R_H = R_g (1 + \sqrt{1 - a^2})$ the radius of the event horizon, $R_g = G M_{BH} / c^2$ the gravitational radius, $M_{BH}$ the mass of the black hole. We introduce the coefficient $k = P_{magn} / P_{acc}$, which lies within $ 0 < k \leq 1 $. Thus, the magnetic field on the event horizon can be expressed as:

\begin{equation}
 B_H = B(R_H) = \frac{\sqrt{2k\dot{M}c}}{R_H},
 \label{eq01}
\end{equation}

In turn, the accretion rate of $\dot{M}$ can be expressed as follows \citep{trakhtenbrot17}:

\begin{equation}
 \dot{M} \simeq 2.4 \left(\frac{\lambda L_{\lambda,45}}{\cos{(i)}}\right)^{3/2} \left(\frac{\lambda}{5100 \textup{\AA}}\right)^2 M_8^{-1} M_{\odot} \text{year}^{-1},
 \label{eq02}
\end{equation}

\noindent where $\lambda$ is the wavelength at which the continuum is measured, $\lambda L_{\lambda, 45} \equiv \lambda L_{\lambda} / 10^{45} erg/s $ is the monochromatic luminosity, $M_8 \equiv M_{BH} / 10^8 M_{\odot}$, $M_{\odot}$ is the mass of the Sun.

Taking $\lambda = 5100 \textup{\AA}$ and substituting (\ref{eq02}) into (\ref{eq01}), we obtain:

\begin{equation}
 B_H = \frac{2.04 \times 10^5 \sqrt{k}}{(1 + \sqrt{1 - a^2}) (\cos{(i)})^{3/4} M_8^{3/2}} \left(\frac{L_{5100}}{10^{45} \text{erg/s}}\right)^{3/4} G.
 \label{eq03}
\end{equation}

The paper \citet{vestergaard06} provides the following method of determination of the mass of SMBH by AGN spectral characteristics:

\begin{equation}
 \frac{M_{BH}}{M_{\odot}} = 10^{6.91} \left(\frac{FWHM(H_\beta)}{10^3 \text{km/s}}\right)^2 \left(\frac{L_{5100}}{10^{44} \text{erg/s}}\right)^{0.5},
 \label{eq04}
\end{equation}

\noindent where $FWHM(H_\beta)$ (hereinafter simply $FWHM$) is the full width at the half amplitude of the emission line $H_\beta$.

Substituting (\ref{eq04}) into (\ref{eq03}) we obtain:

\begin{equation}
 B_H = \frac{1.57 \times 10^6 \sqrt{k}}{(\cos{(i)})^{3/4} (1 + \sqrt{1 - a^2})} \left(\frac{10^3 \text{km/s}}{FWHM}\right)^3 \text{G}.
 \label{eq05}
\end{equation}

In this paper we assume that observed polarization of AGN radiation is generated mostly due to the inclination angle $i$. The parameter $\cos{(i)}$ (also referred to in the literature as $\mu$) is associated with the observed polarization $P$ of AGN radiation. Using the Sobolev-Chandrasekhar theory \citep{sobolev63, chandrasekhar50}, one can calculate the degree of polarization for all $\cos{(i)}$ (see, for example, \citet{gnedin15}). Thus, we obtain the magnetic field strength on the event horizon $B_H$ as a function of the parameters $P$, $k$, $a$ and $FWHM$.

One of the components of the unified AGN model is relativistic jets and/or intense outflows of matter from the surface of the accretion disk. This view is confirmed by an ever-increasing amount of observational data. In addition to jet emissions on a kiloparsec scale, interferometric observations with extra-long bases (VLBI) make it possible to discover and study jet-like structures on a parsec scale. To date, evidence has been obtained for the presence of jets in almost all classes of AGN, including those where they were not previously detected. In addition to the image itself, a powerful indicator of the presence of a relativistic jet is strong synchrotron radio emission. The intensity of the radio emission is related to the kinetic power of the jet $L_j$, which should depend on the spin of SMBH, according to modern theoretical concepts. This circumstance was used in \citet{mikhailov15}, where the following expression is given:

\begin{equation}
 \frac{|a|}{\sqrt{\varepsilon (a)}(1 + \sqrt{1 - a^2})} = \frac{1.81 \eta}{\sqrt{k}} \sqrt{\frac{L_j}{L_{bol}}},
 \label{eq06}
\end{equation}

\noindent where $\varepsilon$ is the coefficient of radiation efficiency of the accretion flux, the coefficient $\eta$ equals $\sqrt{5}$ in the Blandford-Znayek model \citep{blandford77} and $1.05^{-0.5}$ in the hybrid Meier model \citep{meier99}, which combines the mechanisms of Blandford-Znayek and Blandford-Payne \citep{blandford82}. $L_j$ is the kinetic power of the jet, $L_{bol}$ is the bolometric luminosity. In this study we will use the Meier model and take $\eta = 1.05^{-0.5}$.

The objects studied in \citet{afanasiev18} are mainly type I Seyfert galaxies. According to the unified model, type I AGNs have inclination angles that do not exceed $60^{\circ}$, which is in good agreement with the determination of the inclination angles based on polarimetric observations \citep{afanasiev18}. We studied the radiomorphology of the sample based on the NVSS \citep{condon98} survey data. The results are as follows: out of 36 sources, 14 display a radio core (in some cases there are protrusions resembling the base of a jet discharge), 6 sources have a structure of a radio core + one-sided emission, 1 source displays a structure of a radio core + two-sided emission, 6 of the brightest radio sources have a complex structure, 4 weak radio sources display a vague structure and the other 5 radio emission sources were not found on NVSS maps. Objects of type 1 at radio wavelength primarily demonstrate a structure of either the type of a radio core or, in addition, a one-sided jet; out of 36 such are 26 objects (taking into account bright sources with a complex structure, but excluding weak sources). We also note that only one object shows the presence of two-sided emission. Thus, the sample objects show signs of the presence of jets, which allows us to apply to these objects the ratio (\ref{eq06}).

The relationship between the power of the jet and the bolometric luminosity of the disk was studied in a number of works \citep{merloni07, foschini11, daly16}. As a result, empirical relationships were found with coefficients that are consistent within the margin of error. In particular, in \citet{daly16} (see the third row of Table 2 from this work), the following empirical relation was obtained for objects of the type under study:

\begin{equation}
 \log{\frac{L_j}{L_{Edd}}} = (0.41 \pm 0.04) \log{\frac{L_{bol}}{L_{Edd}}} - (1.34 \pm 0.14),
 \label{eq07}
\end{equation}

\noindent where $L_{Edd}$ is the Eddington luminosity. This expression can be converted to:

\begin{equation}
 \frac{L_j}{L_{bol}} = l_E^{(-0.59 \pm 0.04)} \times 10^{(-1.34 \pm 0.14)},
 \label{eq08}
\end{equation}

\noindent where $l_E = L_{bol} / L_{Edd}$ is the Eddington ratio. Substituting (\ref{eq08}) into (\ref{eq06}) we obtain the expression for the parameter $k$:

\begin{equation}
 k = 3.28 \times 10^{(-1.34 \pm 0.14)} \frac{\eta^2 \varepsilon (1 + \sqrt{1 - a^2})^2}{l_E^{(0.59 \pm 0.04)} a^2}.
 \label{eq09}
\end{equation}

Thus, we obtain the final expression for the magnetic field on the event horizon $B_H$:

\begin{equation}
 B_H = \frac{10^{(5.78 \pm 0.07)} \eta \sqrt{\varepsilon}}{l_E^{(0.295 \pm 0.020)} |a| (\cos{(i)})^{3/4}} \left(\frac{10^3 \text{km/s}}{FWHM}\right)^3 \text{G}.
 \label{eq10}
\end{equation}

It should be noted that for spin values of $0.5 < a < 0.998$ specific for most objects of this type, the dependence of $B_H$ on spin is rather small (taking into account the fact that $\varepsilon$ also depends on spin \citep{bardeen72}). In particular, the maximum value of the magnetic field, which is achieved at $ a = 0.5 $, is greater than the minimum at $a \approx 0.85$ by the factor of only about 1.3. Thus, even a large error in determination of the spin will not affect the determination of $B_H$ significantly.

To determine the Eddington ratio $l_E = L_{bol} / L_{Edd}$ we use the following expression: $ L_{Edd} = 1.5 \times 10^{38} M_{BH} / M_{\odot}$ erg/s, $L_{bol} = L_{5100} \times BC$, where $BC$ is the bolometric correction, which we calculate using the formula \citep{afanasiev18}

\begin{equation}
 \log(BC) = -0.54 \log(M_{BH}/M_{\odot}) + 5.43.
 \label{eq11}
\end{equation}

The radiation efficiency coefficient $\varepsilon(a)$ is determined by the formula \citep{du14}

\begin{equation}
 (\cos{(i)})^{3/2} l_E = 0.201 \left(\frac{L_{5100}}{10^{44} erg/s}\right)^{3/2} \frac{\varepsilon (a)}{M_8^2}.
 \label{eq12}
\end{equation}

In turn, the spin $a$ is determined numerically using the relation \citep{bardeen72}

\begin{equation}
 \varepsilon(a) = 1 - \frac{R_{ISCO}^{3/2} - 2 R_{ISCO}^{1/2} + |a|}{R_{ISCO}^{3/4}(R_{ISCO}^{3/2} - 3 R_{ISCO}^{1/2} + 2 |a|)^{1/2}},
 \label{eq13}
\end{equation}

\noindent where $R_{ISCO}$ is the radius of the innermost stable circular orbit of a black hole, which is expressed through the spin as follows:

\begin{equation}
 \begin{array}{l}
  R_{ISCO}(a) = 3 + Z_2 \pm ((3 - Z_1)(3 + Z_1 + 2 Z_2))^{1/2},\\
  Z_1 = 1 + (1 - a^2)^{1/3}((1 + a)^{1/3} + (1 - a)^{1/3}),\\
  Z_2 = (3 a^2 + Z_1^2)^{1/2}.
 \end{array}
 \label{eq14}
\end{equation}

\noindent In the expression for $R_{ISCO}(a)$, the sign ''-'' is used for prograde ($a \geq 0$), and the sign ''+'' for retrograde rotation ($a < 0$).

\section{Results of the calculations} 

\subsection{Theoretical curves} 

\begin{figure}
	\centering
  \includegraphics[width=\columnwidth,trim= 35 34 52 50, clip=true]{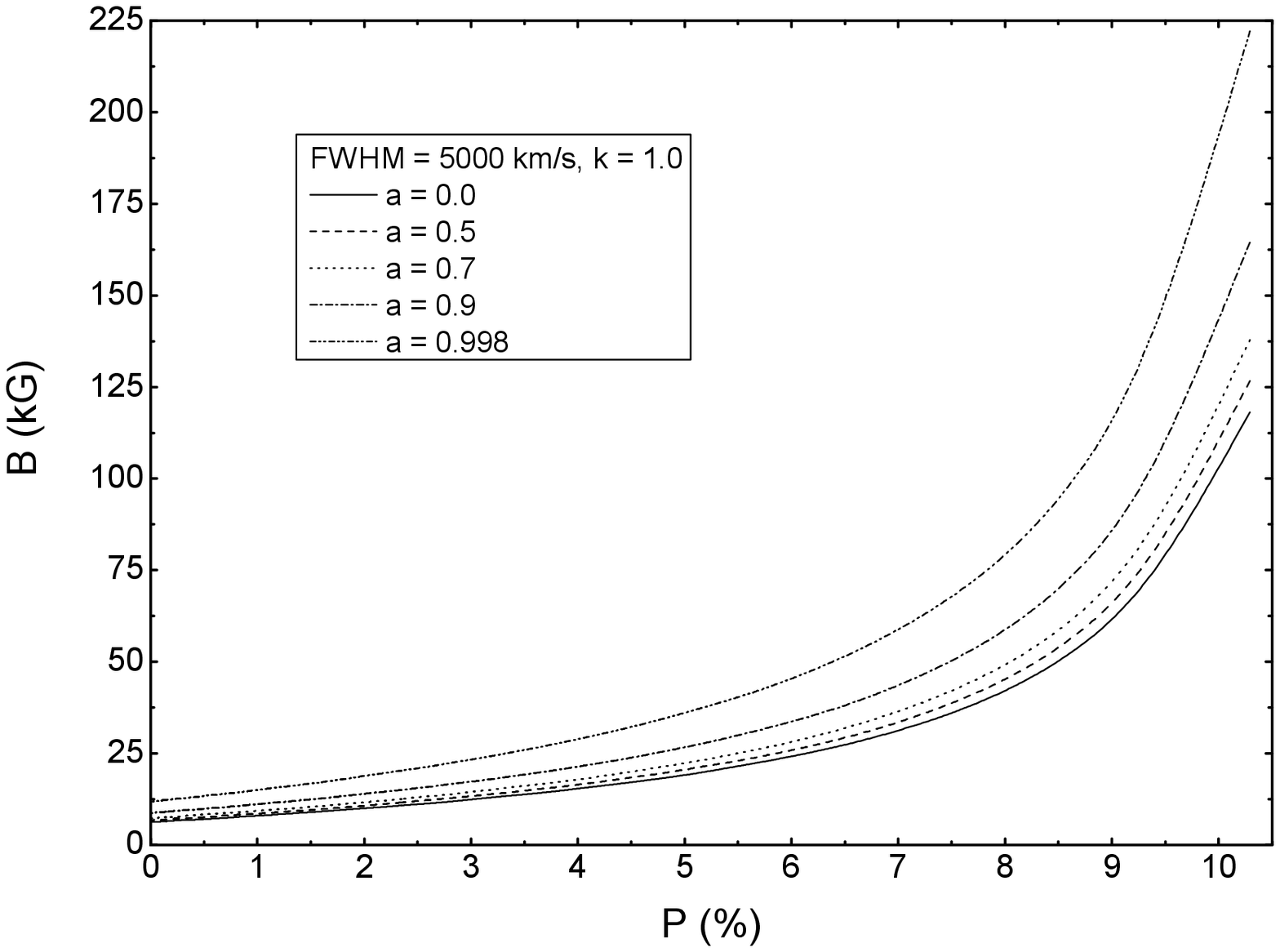}
  \caption{Magnetic field $B$ versus polarization $P$ for different values of spin $a$ and fixed $FWHM$ and $k$.}
  \label{fig01}
\end{figure}

\begin{figure}
	\centering
  \includegraphics[width=\columnwidth,trim= 35 34 52 48, clip=true]{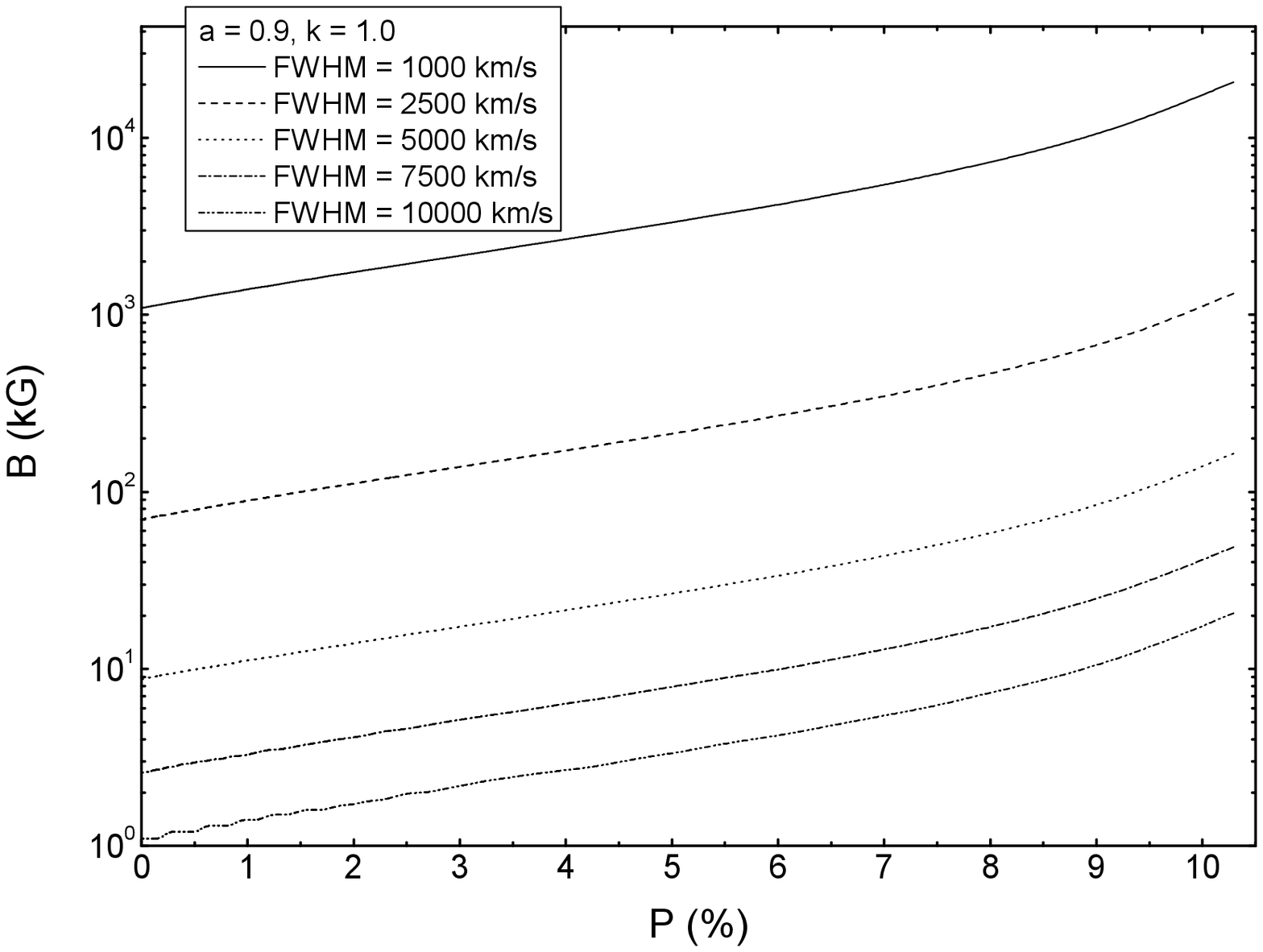}
  \caption{The dependence of the magnetic field $B$ on the polarization $P$ for different $FWHM$ and fixed $a$ and $k$.}
  \label{fig02}
\end{figure}

\begin{figure}
	\centering
  \includegraphics[width=\columnwidth,trim= 52 33 74 40, clip=true]{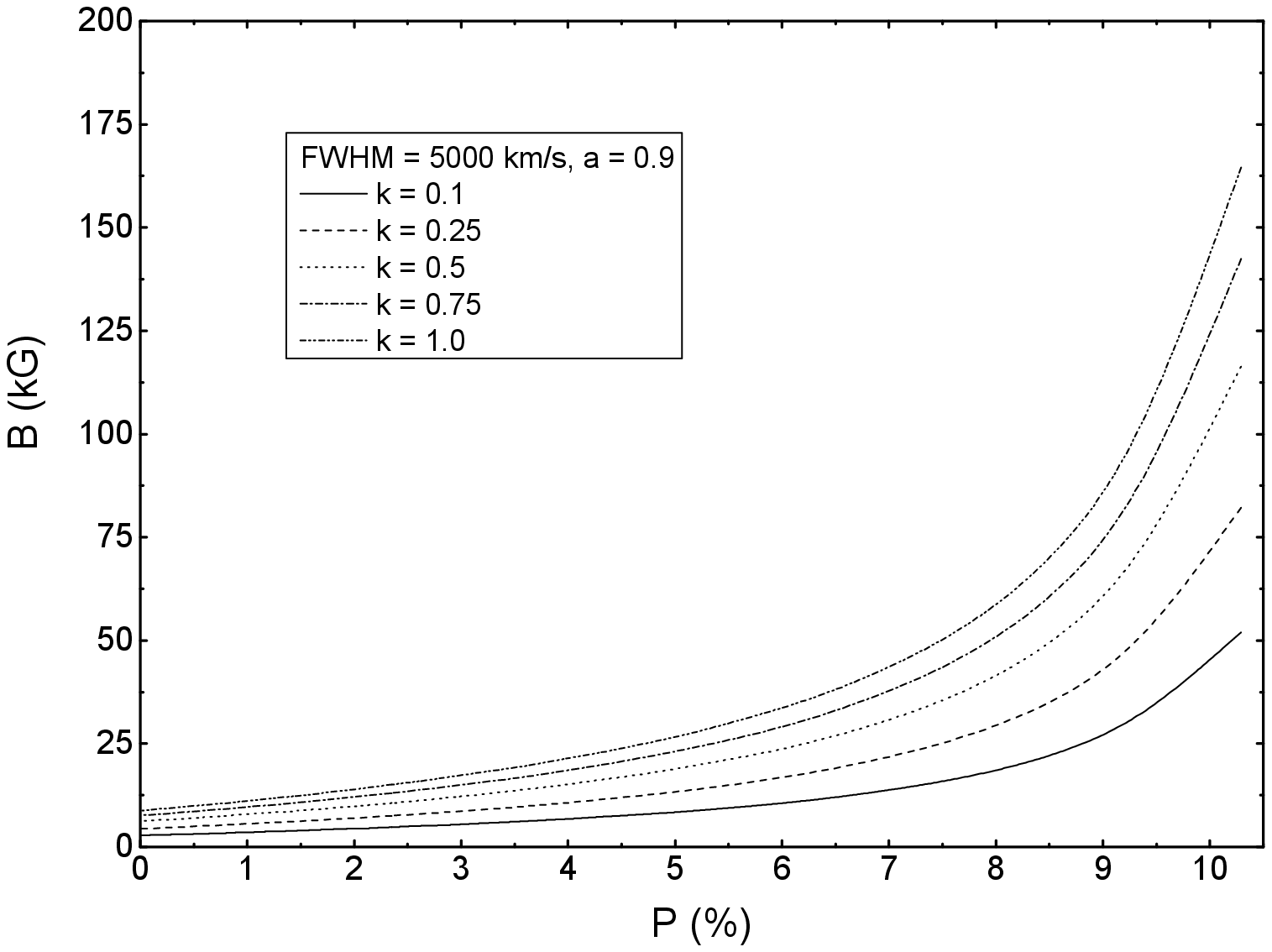}
  \caption{Magnetic field $B$ versus polarization $P$ for different values of the parameter $k$ and fixed $FWHM$ and $a$.}
  \label{fig03}
\end{figure}

Figures \ref{fig01}, \ref{fig02} and \ref{fig03} show graphs of $B_H$ versus $P$ for different values of the parameters $k$, $a$ and $FWHM$. Fig. \ref{fig01} demonstrates that the dependence of the curves on the spin is not particularly strong, especially taking into account the fact that the polarization of the observed objects, as a rule, does not exceed 2\%, and the spin exceeds 0.5. In Fig.\ref{fig02} we can see that the magnetic field strongly depends on the parameter $FWHM$, and, in particular, at $FHMW \approx 10^4$ km/s (which is close to $FWHM$ for some observed objects) the magnetic field strength can reach the order of one megagauss even with the observed polarization of about 1-2\%. Fig.\ref{fig03} indicates that the magnetic field is noticeably dependent on the parameter $k$. For $k = 0.1$ and $k = 1.0$, the magnetic field strength differs by more than 3 times.

\subsection{Calculation of magnetic fields using data from the literature} 

\renewcommand{\arraystretch}{1.5}
\begin{table*}
\label{tab01}
\footnotesize
\caption{Results of determination of the magnetic field strength on the event horizon for our objects. FWHM is expressed in km/s, $B_H$ in gauss, $i$ in degrees, $L_ {5100}$ in erg/s.}
\setlength{\tabcolsep}{4pt}
\centering
\begin{tabular}{|l|c|c|l|c|c|c|c|c|c|}
\hline
Object & $\log\left(\frac{M_{BH}}{M_\odot}\right)$ & $\log(L_{5100})$ & FWHM & $i$ & $a$ & $l_E$ & $\varepsilon$ & $\log(k)$ & $\log(B_H)$ \\
\hline
PG 0003+199 & $7.42^{+0.02}_{-0.02}$ & $43.67^2$ & $2182 \pm 53^2$ & $31^{+4}_{-5}$ & $0.990^{+0.008}_{-0.024}$ & $0.308^{+0.024}_{-0.022}$ & $0.27^{+0.05}_{-0.06}$ & $-0.99^{+0.32}_{- 0.31}$ & $4.68^{+0.17}_{-0.17}$\\
PG 0007+106 & $8.14^{+0.01}_{-0.01}$ & $44.29^2$ & $3494 \pm 35^6$ & $31^{+5}_{-7}$ & $0.993^{+0.005}_{-0.011}$ & $0.100^{+0.003}_{-0.003}$ & $0.28^{+0.04}_{-0.04}$ & $-0.71^{+0.23}_{- 0.23}$ & $4.21^{+0.12}_{-0.12}$\\
PG 0026+129 & $8.09^{+0.02}_{-0.02}$ & $44.70^2$ & $2598 \pm 57^2$ & $45^{+4}_{-6}$ & $0.806^{+0.113}_{-0.175}$ & $0.311^{+0.022}_{-0.020}$ & $0.12^{+0.04}_{-0.03}$ & $-0.86^{+0.57}_{- 0.48}$ & $4.43^{+0.29}_{-0.24}$\\
PG 0049+171 & $8.35^{+0.04}_{-0.04}$ & $44.00^1$ & $5234 \pm 260^1$ & $38^{+4}_{-5}$ & $0.998^{+0.000}_{-0.006}$ & $0.025^{+0.004}_{-0.003}$ & $0.32^{+0.00}_{-0.05}$ & $-0.34^{+0.17}_{- 0.19}$ & $3.92^{+0.14}_{-0.18}$\\
PG 0054+144 & $8.97^{+0.05}_{-0.05}$ & $44.78^5$ & $6830 \pm 371^4$ & $43^{+7}_{-9}$ & $0.996^{+0.002}_{-0.073}$ & $0.017^{+0.003}_{-0.002}$ & $0.30^{+0.02}_{-0.13}$ & $-0.24^{+0.41}_{- 0.39}$ & $3.64^{+0.22}_{-0.29}$\\
PG 0157+001 & $8.17^{+0.04}_{-0.04}$ & $44.98^1$ & $2431 \pm 120^1$ & $39^{+6}_{-7}$ & $0.756^{+0.185}_{-0.462}$ & $0.441^{+0.074}_{-0.061}$ & $0.11^{+0.07}_{-0.04}$ & $-0.90^{+1.34}_{- 0.76}$ & $4.45^{+0.69}_{-0.38}$\\
PG 0804+761 & $8.22^{+0.01}_{-0.01}$ & $44.61^2$ & $3190 \pm 39^2$ & $44^{+6}_{-9}$ & $0.912^{+0.057}_{-0.098}$ & $0.157^{+0.006}_{-0.006}$ & $0.16^{+0.05}_{-0.04}$ & $-0.78^{+0.44}_{- 0.39}$ & $4.25^{+0.22}_{-0.20}$\\
PG 0844+349 & $7.93^{+0.04}_{-0.04}$ & $44.60^3$ & $2300 \pm 115^3$ & $36^{+4}_{-5}$ & $0.873^{+0.102}_{-0.257}$ & $0.429^{+0.073}_{-0.060}$ & $0.14^{+0.08}_{-0.05}$ & $-1.01^{+0.81}_{- 0.63}$ & $4.50^{+0.41}_{-0.31}$\\
PG 0921+525 & $7.17^{+0.02}_{-0.03}$ & $43.15^2$ & $2194 \pm 64^2$ & $34^{+4}_{-4}$ & $0.998^{+0.000}_{-0.008}$ & $0.230^{+0.022}_{-0.020}$ & $0.32^{+0.00}_{-0.06}$ & $-0.91^{+0.20}_{- 0.22}$ & $4.75^{+0.13}_{-0.16}$\\
PG 0923+129 & $7.25^{+0.04}_{-0.04}$ & $43.26^3$ & $2270 \pm 110^3$ & $18^{+6}_{-8}$ & $0.998^{+0.000}_{-0.005}$ & $0.220^{+0.036}_{-0.030}$ & $0.32^{+0.00}_{-0.04}$ & $-0.90^{+0.20}_{- 0.22}$ & $4.67^{+0.15}_{-0.18}$\\
PG 0923+201 & $9.02^{+0.04}_{-0.04}$ & $45.06^3$ & $6140 \pm 300^3$ & $28^{+6}_{-8}$ & $0.996^{+0.002}_{-0.041}$ & $0.027^{+0.004}_{-0.004}$ & $0.30^{+0.02}_{-0.11}$ & $-0.35^{+0.32}_{- 0.33}$ & $3.66^{+0.17}_{-0.24}$\\
PG 0953+414 & $8.59^{+0.01}_{-0.01}$ & $45.36^2$ & $3155 \pm 44^2$ & $29^{+3}_{-4}$ & $0.847^{+0.055}_{-0.072}$ & $0.242^{+0.011}_{-0.010}$ & $0.14^{+0.02}_{-0.02}$ & $-0.84^{+0.32}_{- 0.30}$ & $4.14^{+0.16}_{-0.15}$\\
PG 1022+519 & $7.12^{+0.01}_{-0.01}$ & $43.72^2$ & $1489 \pm 15^2$ & $41^{+5}_{-7}$ & $0.888^{+0.057}_{-0.082}$ & $1.027^{+0.033}_{-0.031}$ & $0.15^{+0.04}_{-0.03}$ & $-1.24^{+0.40}_{- 0.37}$ & $4.98^{+0.20}_{-0.19}$\\
PG 1116+215 & $8.57^{+0.04}_{-0.04}$ & $45.13^3$ & $3530 \pm 175^3$ & $34^{+3}_{-4}$ & $0.907^{+0.076}_{-0.192}$ & $0.152^{+0.026}_{-0.021}$ & $0.16^{+0.08}_{-0.05}$ & $-0.77^{+0.69}_{- 0.56}$ & $4.08^{+0.34}_{-0.28}$\\
PG 1309+355 & $9.06^{+0.05}_{-0.06}$ & $45.08^3$ & $6370 \pm 410^4$ & $36^{+10}_{-20}$ & $0.991^{+0.007}_{-0.137}$ & $0.024^{+0.005}_{-0.004}$ & $0.27^{+0.05}_{-0.14}$ & $-0.34^{+0.58}_{- 0.49}$ & $3.63^{+0.30}_{-0.36}$\\
PG 1351+695 & $8.27^{+0.02}_{-0.02}$ & $43.85^2$ & $5208 \pm 95^2$ & $34^{+4}_{-6}$ & $0.998^{+0.000}_{-0.000}$ & $0.023^{+0.001}_{-0.001}$ & $0.32^{+0.00}_{-0.00}$ & $-0.32^{+0.09}_{- 0.09}$ & $3.92^{+0.08}_{-0.09}$\\
PG 1354+213 & $8.63^{+0.04}_{-0.04}$ & $44.98^1$ & $4126 \pm 200^1$ & $47^{+6}_{-8}$ & $0.889^{+0.099}_{-0.315}$ & $0.087^{+0.014}_{-0.012}$ & $0.15^{+0.11}_{-0.06}$ & $-0.61^{+0.94}_{- 0.66}$ & $4.01^{+0.48}_{-0.34}$\\
PG 1425+267 & $9.74^{+0.04}_{-0.04}$ & $45.76^1$ & $9404 \pm 470^1$ & $53^{+5}_{-7}$ & $0.915^{+0.079}_{-0.281}$ & $0.010^{+0.002}_{-0.001}$ & $0.16^{+0.13}_{-0.07}$ & $-0.09^{+0.87}_{- 0.62}$ & $3.25^{+0.44}_{-0.32}$\\
PG 1434+590 & $8.30^{+0.02}_{-0.02}$ & $44.00^2$ & $4937 \pm 120^2$ & $38^{+3}_{-4}$ & $0.998^{+0.000}_{-0.001}$ & $0.030^{+0.002}_{-0.002}$ & $0.32^{+0.00}_{-0.00}$ & $-0.38^{+0.11}_{- 0.10}$ & $3.98^{+0.09}_{-0.10}$\\
PG 1501+106 & $8.53^{+0.04}_{-0.04}$ & $44.28^1$ & $5454 \pm 270^1$ & $41^{+6}_{-8}$ & $0.998^{+0.000}_{-0.031}$ & $0.025^{+0.004}_{-0.004}$ & $0.32^{+0.00}_{-0.11}$ & $-0.34^{+0.28}_{- 0.30}$ & $3.88^{+0.17}_{-0.25}$\\
PG 1545+210 & $9.32^{+0.04}_{-0.04}$ & $45.43^1$ & $7021 \pm 350^1$ & $51^{+5}_{-6}$ & $0.916^{+0.077}_{-0.266}$ & $0.021^{+0.004}_{-0.003}$ & $0.16^{+0.11}_{-0.07}$ & $-0.27^{+0.83}_{- 0.61}$ & $3.52^{+0.42}_{-0.31}$\\
PG 1613+658 & $9.18^{+0.03}_{-0.03}$ & $44.70^2$ & $9142 \pm 288^2$ & $44^{+4}_{-5}$ & $0.998^{+0.000}_{-0.004}$ & $0.006^{+0.001}_{-0.001}$ & $0.32^{+0.00}_{-0.03}$ & $0.01^{+0.12}_{- 0.12}$ & $3.40^{+0.10}_{-0.13}$\\
PG 1700+518 & $7.74^{+0.02}_{-0.02}$ & $44.27^2$ & $2230 \pm 57^2$ & $40^{+4}_{-5}$ & $0.905^{+0.061}_{-0.114}$ & $0.396^{+0.033}_{-0.030}$ & $0.16^{+0.05}_{-0.04}$ & $-1.01^{+0.49}_{- 0.43}$ & $4.58^{+0.25}_{-0.22}$\\
PG 1704+608 & $9.39^{+0.04}_{-0.04}$ & $45.70^1$ & $6552 \pm 330^1$ & $38^{+0}_{-0}$ & $0.942^{+0.047}_{-0.120}$ & $0.031^{+0.005}_{-0.004}$ & $0.18^{+0.08}_{-0.06}$ & $-0.39^{+0.53}_{- 0.45}$ & $3.51^{+0.26}_{-0.22}$\\
PG 2112+059 & $8.61^{+0.04}_{-0.04}$ & $45.48^9$ & $3010 \pm 150^9$ & $44^{+4}_{-4}$ & $0.583^{+0.278}_{-0.501}$ & $0.299^{+0.051}_{-0.042}$ & $0.09^{+0.05}_{-0.03}$ & $-0.60^{+2.14}_{- 0.83}$ & $4.31^{+1.12}_{-0.42}$\\
PG 2130+099 & $7.58^{+0.00}_{-0.00}$ & $44.33^2$ & $1781 \pm 5^6$ & $35^{+4}_{-4}$ & $0.854^{+0.033}_{-0.046}$ & $0.817^{+0.007}_{-0.007}$ & $0.14^{+0.01}_{-0.01}$ & $-1.16^{+0.26}_{- 0.25}$ & $4.75^{+0.13}_{-0.13}$\\
PG 2209+184 & $8.41^{+0.04}_{-0.04}$ & $43.69^3$ & $6690 \pm 330^3$ & $41^{+5}_{-7}$ & $0.998^{+0.000}_{-0.000}$ & $0.010^{+0.002}_{-0.001}$ & $0.32^{+0.00}_{-0.00}$ & $-0.10^{+0.10}_{- 0.10}$ & $3.73^{+0.14}_{-0.14}$\\
PG 2214+139 & $8.55^{+0.04}_{-0.04}$ & $44.66^1$ & $4532 \pm 226^1$ & $36^{+5}_{-6}$ & $0.988^{+0.010}_{-0.074}$ & $0.055^{+0.009}_{-0.008}$ & $0.26^{+0.06}_{-0.10}$ & $-0.55^{+0.46}_{- 0.41}$ & $3.96^{+0.23}_{-0.26}$\\
PG 2251+113 & $8.99^{+0.04}_{-0.04}$ & $45.69^1$ & $4147 \pm 207^1$ & $41^{+5}_{-6}$ & $0.757^{+0.184}_{-0.443}$ & $0.124^{+0.021}_{-0.017}$ & $0.11^{+0.07}_{-0.04}$ & $-0.58^{+1.25}_{- 0.73}$ & $3.93^{+0.65}_{-0.36}$\\
PG 2304+042 & $8.55^{+0.04}_{-0.04}$ & $43.59^3$ & $8390 \pm 419^3$ & $49^{+4}_{-5}$ & $0.998^{+0.000}_{-0.000}$ & $0.005^{+0.001}_{-0.001}$ & $0.32^{+0.00}_{-0.00}$ & $0.09^{+0.08}_{- 0.09}$ & $3.58^{+0.14}_{-0.14}$\\
PG 2308+098 & $9.60^{+0.04}_{-0.04}$ & $45.78^1$ & $7914 \pm 395^1$ & $48^{+5}_{-6}$ & $0.912^{+0.079}_{-0.260}$ & $0.018^{+0.003}_{-0.002}$ & $0.16^{+0.11}_{-0.07}$ & $-0.22^{+0.81}_{- 0.60}$ & $3.37^{+0.41}_{-0.30}$\\
Mrk 1146 & $7.41^{+0.04}_{-0.04}$ & $43.28^3$ & $2700 \pm 135^3$ & $44^{+7}_{-9}$ & $0.997^{+0.001}_{-0.056}$ & $0.130^{+0.022}_{-0.018}$ & $0.32^{+0.00}_{-0.14}$ & $-0.75^{+0.38}_{- 0.40}$ & $4.60^{+0.21}_{-0.30}$\\
LEDA 3095776 & $7.12^{+0.04}_{-0.04}$ & $43.60^5$ & $1600 \pm 80^5$ & $39^{+6}_{-8}$ & $0.939^{+0.056}_{-0.199}$ & $0.773^{+0.132}_{-0.108}$ & $0.18^{+0.12}_{-0.07}$ & $-1.21^{+0.79}_{- 0.62}$ & $4.93^{+0.40}_{-0.32}$\\
LEDA 2325569 & $6.58^{+0.04}_{-0.04}$ & $42.90^5$ & $1290 \pm 64^5$ & $39^{+6}_{-8}$ & $0.977^{+0.021}_{-0.116}$ & $1.035^{+0.176}_{-0.143}$ & $0.23^{+0.09}_{-0.09}$ & $-1.31^{+0.62}_{- 0.53}$ & $5.21^{+0.31}_{-0.30}$\\
3C 390.3 & $9.12^{+0.09}_{-0.10}$ & $44.43^7$ & $9958 \pm 1046^8$ & $55^{+3}_{-2}$ & $0.998^{+0.000}_{-0.102}$ & $0.004^{+0.002}_{-0.001}$ & $0.32^{+0.00}_{-0.17}$ & $0.11^{+0.48}_{- 0.46}$ & $3.41^{+0.28}_{-0.37}$\\
NGC 7469 & $7.84^{+0.02}_{-0.02}$ & $43.78^2$ & $3296 \pm 75^2$ & $22^{+4}_{-4}$ & $0.998^{+0.000}_{-0.000}$ & $0.092^{+0.007}_{-0.006}$ & $0.32^{+0.00}_{-0.00}$ & $-0.67^{+0.12}_{- 0.12}$ & $4.30^{+0.10}_{-0.10}$\\
\hline
\multicolumn{10}{l}{Data sources: (1) \citet{vestergaard06}; (2) \citet{feng14}; (3) \citet{sani10};}\\
\multicolumn{10}{l}{(4) \citet{sheinis17}; (5) \citet{decarli11}; (6) \citet{grier17}; (7) \citet{bentz13};}\\
\multicolumn{10}{l}{(8) \citet{kollatschny18}; (9) \citet{grupe15}; Data on the $i$ from paper \citet{afanasiev18}.}\\
\end{tabular}
\end{table*}
\renewcommand{\arraystretch}{1.0}


Table 1 presents the results of calculations of the mass $M_{BH}$, spin $a$, radiation efficiency coefficient $\varepsilon$, Eddington ratio $l_E$, parameter $k$ and magnetic field $B_H$ at the event horizon for 36 AGNs. For all this objects, the parameter $i$ was previously determined by us from polarimetric observations \citep{afanasiev18}. The values of $FWHM$ and $L_ {5100}$ were taken from \citet{vestergaard06, feng14, sani10, sheinis17, decarli11, grier17, bentz13, kollatschny18, grupe15}. If the determination errors for $FWHM$ were not explicitly indicated, they were accepted to be $\sim 5 \% $.

\begin{figure}
	\centering
  \includegraphics[width=\columnwidth,trim= 60 21 28 38, clip=true]{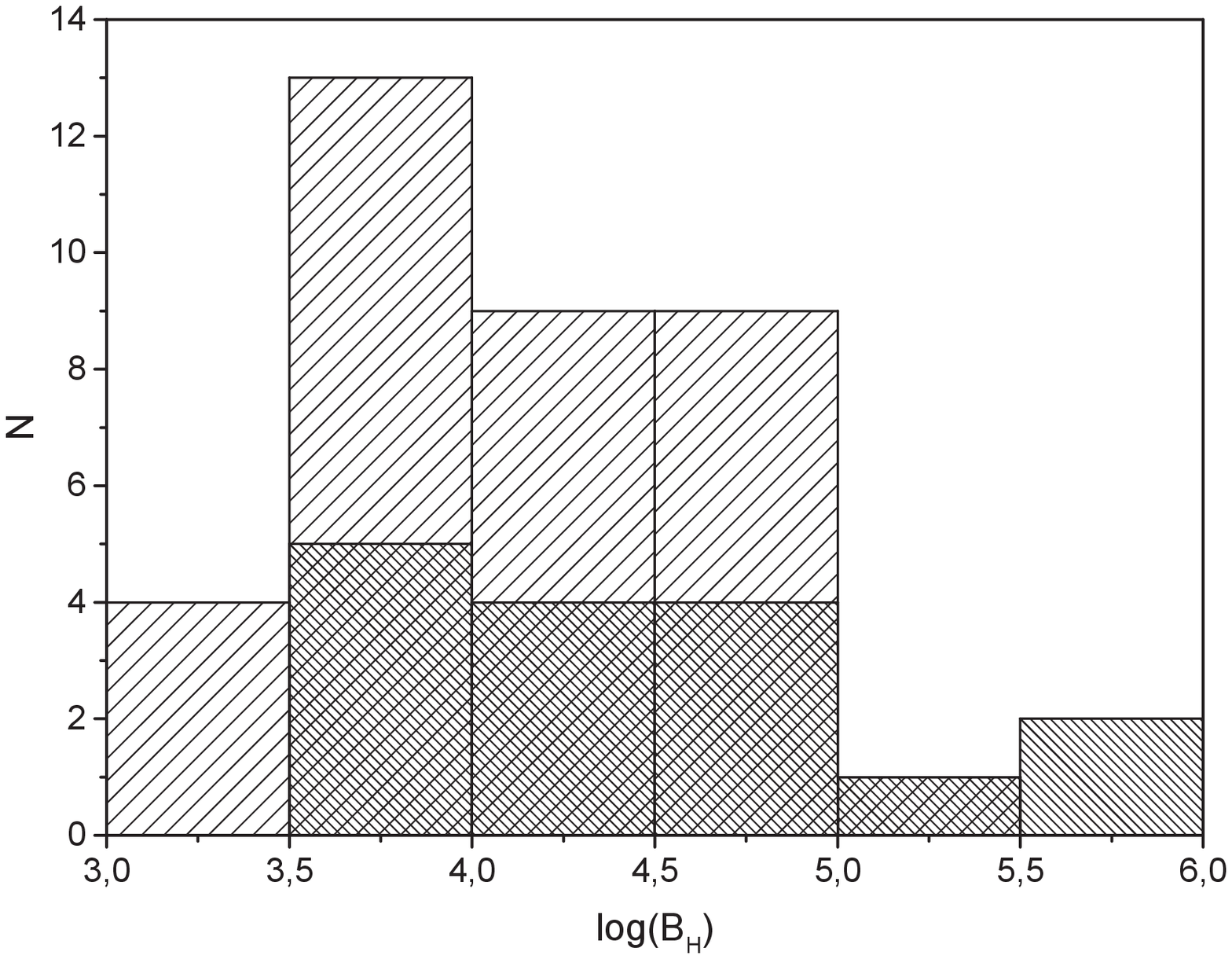}
  \caption{A histogram showing the number of SMBHs in the AGN depending on the magnetic field strength on their event horizon. Objects from sample 1 are shown by sparse hatching, objects from sample 2 by dense hatching.}
  \label{fig04}
\end{figure}

A histogram in Fig.\ref{fig04} presents the number of SMBHs in AGNs depending on the logarithm of the magnetic field (in gauss) on their event horizon. Considered objects are shown by sparse hatching. It can be seen that most of SMBHs have a magnetic field on the event horizon ranging from several to tens of kG, which is consistent with estimates made by other authors for objects of this type \citep{baczko16, mikhailov15, gnedin13b, pariev03, daly19}.

Our results are consistent with those of a numerical simulations of geometrically thin disks with the magnetic field \citep{pariev03}, where the magnetic field strength was obtained for objects of the studied type with a characteristic mass $ M \approx 10^8 M_\odot$ over a radius of $R = 20 R_g$ up to $10^4 - 10^5$G, which corresponds to the fields on the event horizon in the region of $ 5 \times 10^4 - 10^6$G.

Estimates of the magnetic fields of accretion disks for a large number of AGNs were obtained in \citet{daly19}. The average value of the field in this work turned out to be $10^4$G, which fits well with our results. In addition, the list of objects of this work contains 9 objects from our list. For 6 of them (PG~0003+199, PG~0026+129, PG~0804+761, PG~0844+349, PG~0953+414 and PG~2130+099) the obtained magnetic field strength estimates are close within the error range. For the remaining 3 objects (PG~1351+695, PG~1613+658 and NGC~7469), the values of the magnetic field strength in \citet{daly19} turned out to be 5-10 times larger, which may be due to differences in the method of determination of the bolometric luminosity and masses of SMBHs. In particular, we determined the bolometric luminosity through the luminosity at 5100 angstroms \citep{afanasiev18}, and in the paper \citet{daly19} this is done through X-ray luminosity.

\begin{figure}
	\centering
  \includegraphics[width=\columnwidth,trim= 62 26 62 46, clip=true]{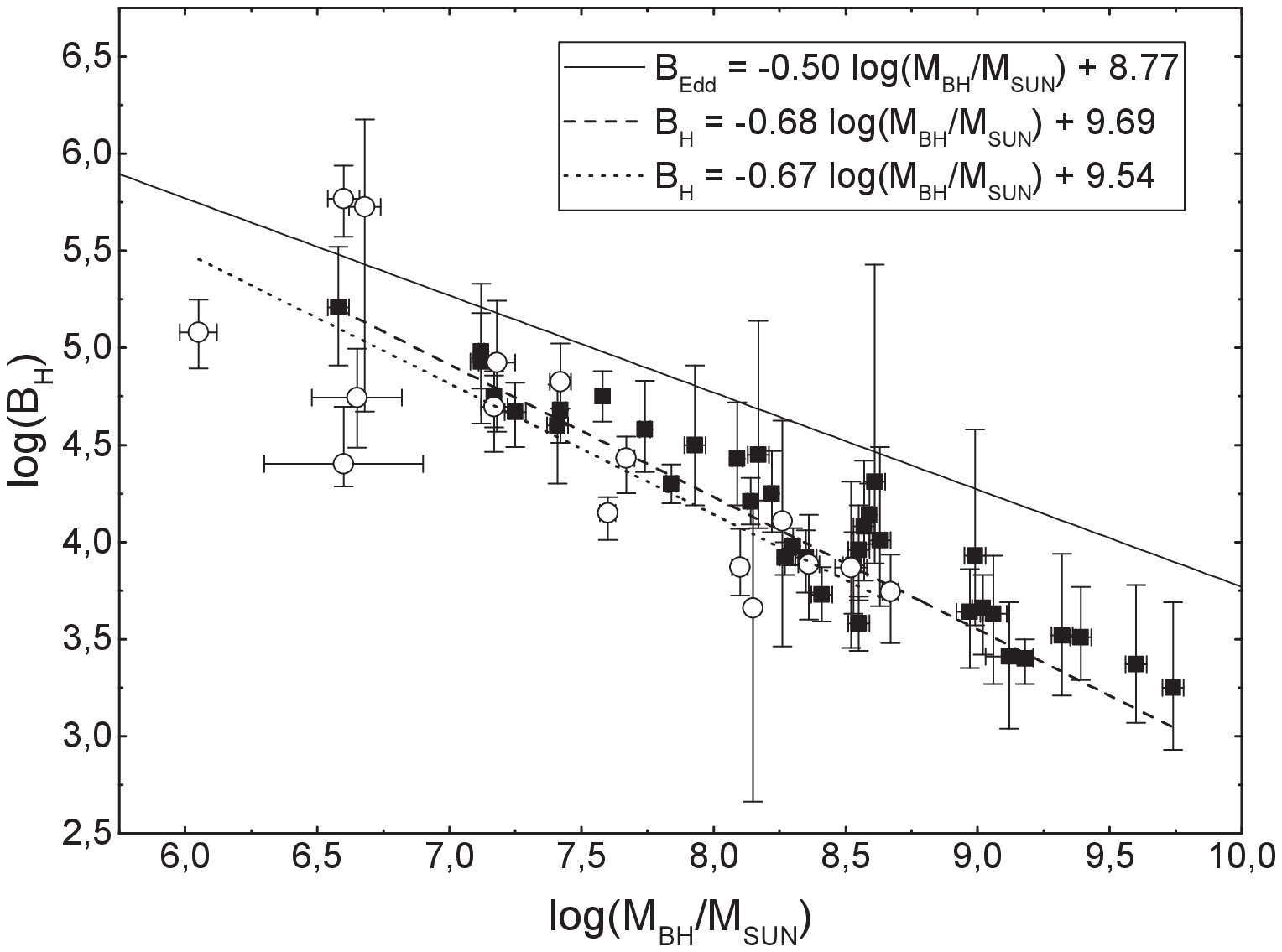}
  \caption{Dependence of the magnetic field strength on the event horizon on the mass of SMBH. Black squares denote objects from sample 1, white circles from sample 2.}
  \label{fig05}
\end{figure}

\begin{figure}
	\centering
  \includegraphics[width=\columnwidth,trim= 60 26 72 46, clip=true]{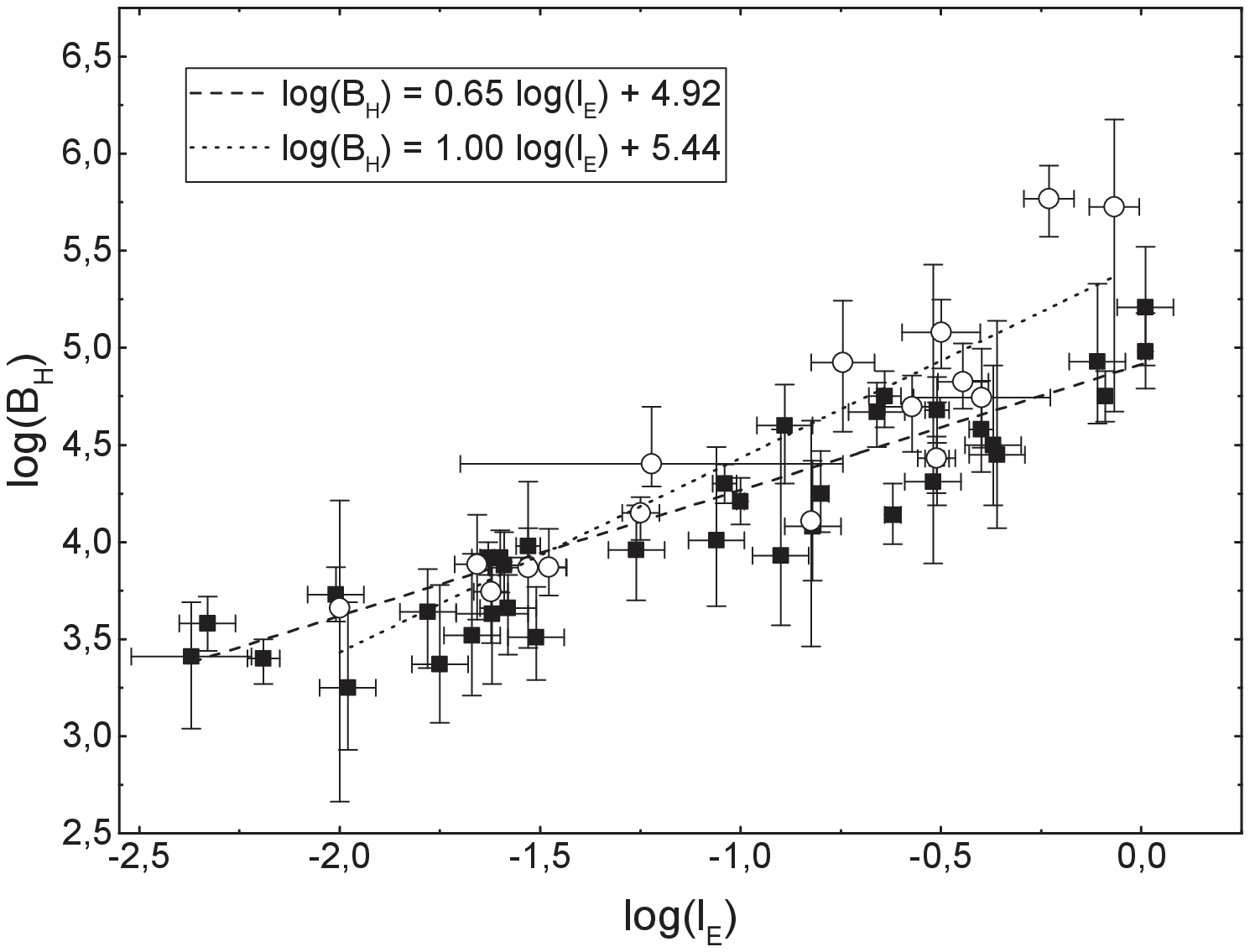}
  \caption{Dependence of the magnetic field strength on the event horizon on the Eddington ratio. Black squares denote objects from sample 1, white circles from sample 2.}
  \label{fig06}
\end{figure}

\begin{figure}
	\centering
  \includegraphics[width=\columnwidth,trim= 60 26 72 46, clip=true]{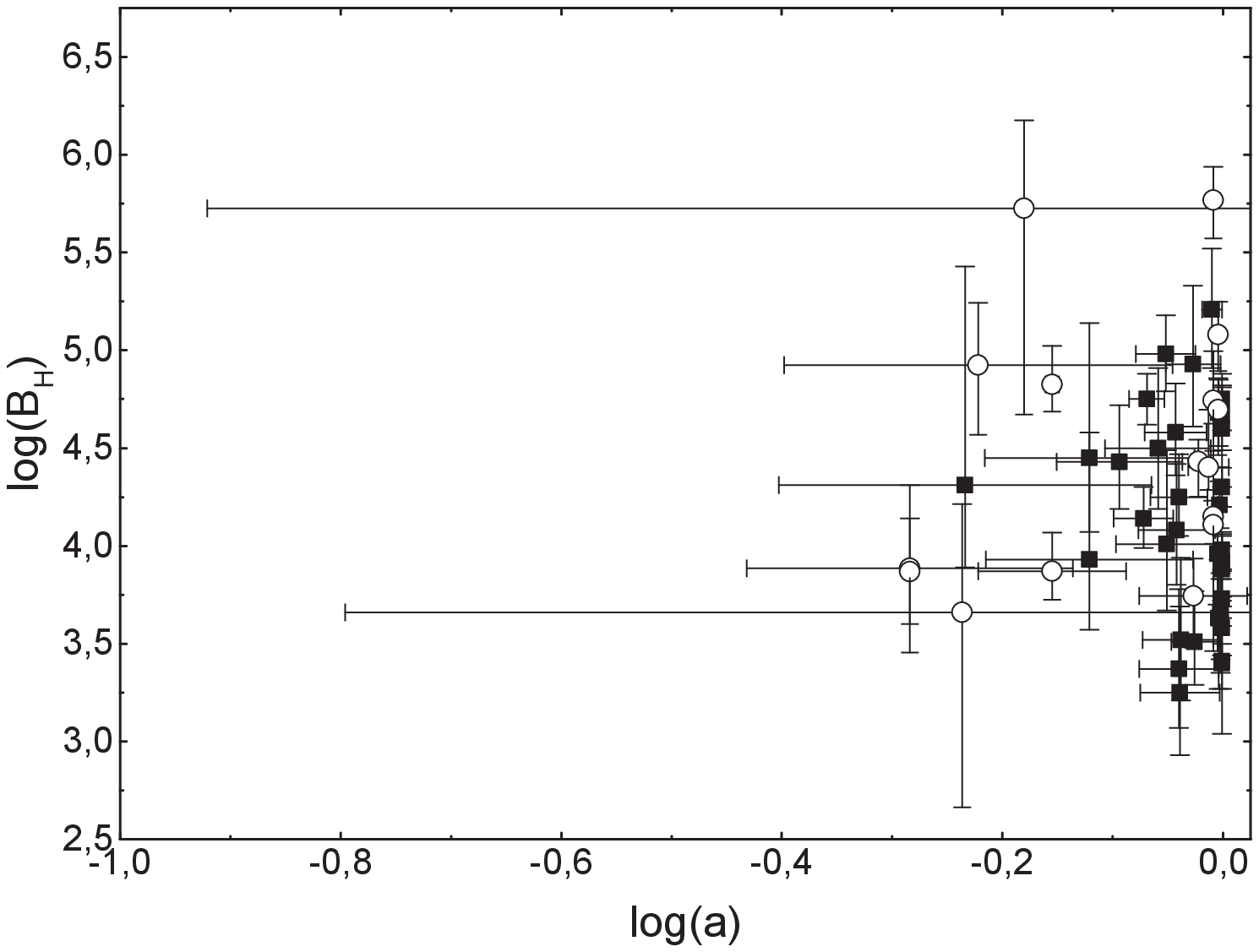}
  \caption{Dependence of the magnetic field strength on the event horizon on the spin of SMBH. Black squares denote objects from sample 1, white circles from sample 2.}
  \label{fig07}
\end{figure}

\begin{figure}
	\centering
  \includegraphics[width=\columnwidth,trim= 60 26 72 46, clip=true]{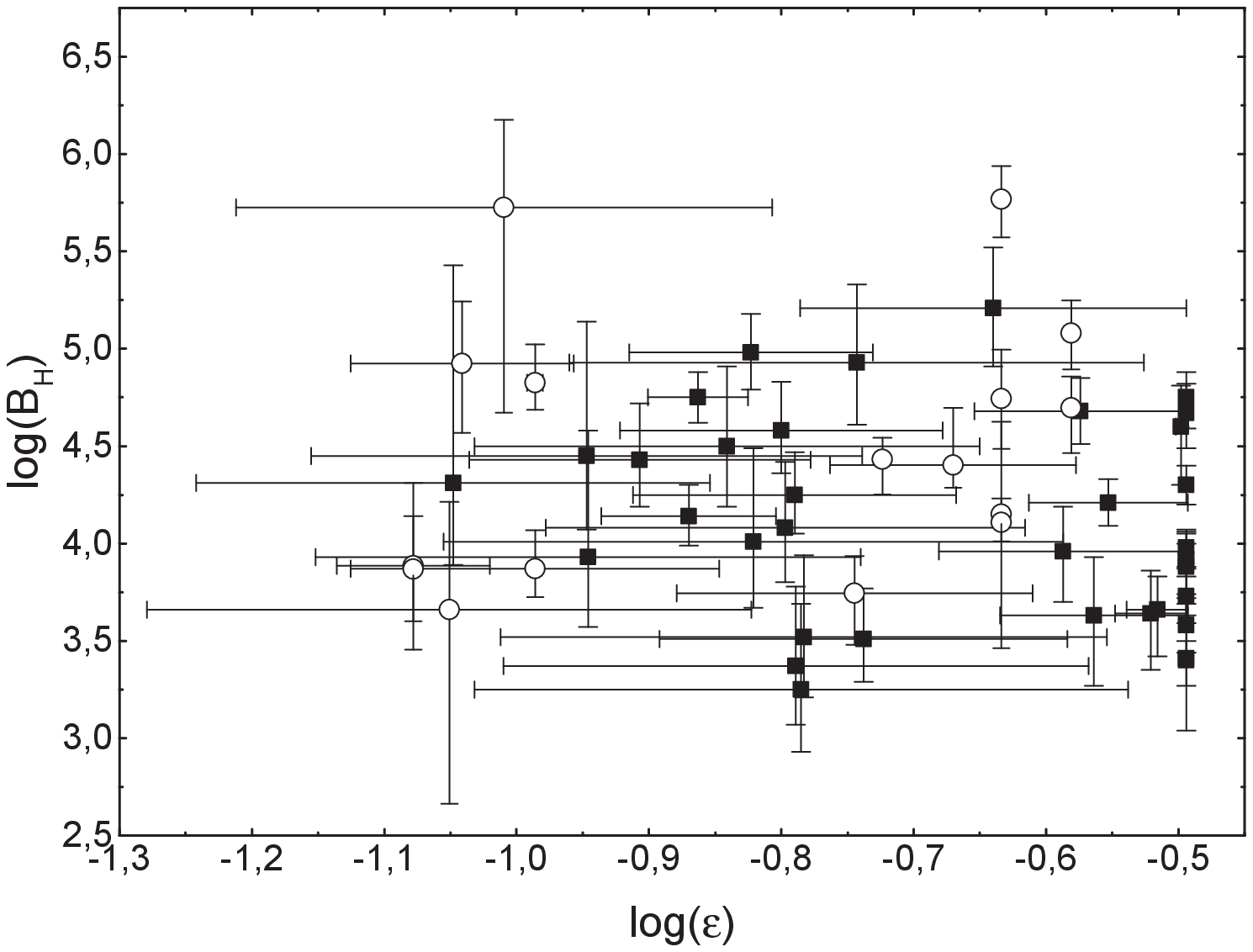}
  \caption{Dependence of the magnetic field strength on the event horizon on the coefficient of radiation efficiency of SMBH. Black squares denote objects from sample 1, white circles from sample 2.}
  \label{fig08}
\end{figure}

In Figs.\ref{fig05}-\ref{fig08} the considered objects are shown by black squares.

Fig.\ref{fig05} shows the dependence of the magnetic field strength on the event horizon on the mass of the SMBH. The field depends on the mass inversely: $\log(B_H) \approx - (0.68 \pm 0.04) \log(M_{BH} / M_{\odot}) + (9.69 \pm 0.32)$, which is consistent with the mass dependence of the Eddington magnetic field of the accretion disk: $\log(B_{Edd}) \approx - 0.5 \log(M_{BH} / M_{\odot}) + 8.77 $ \citep {daly19} .

Fig.\ref{fig06} shows the dependence of the magnetic field strength on the event horizon on the Eddington ratio. A dependence of the form $\log(B_H) \approx (0.65 \pm 0.04) \log(l_E) + (4.92 \pm 0.06)$ is present.

Figures \ref{fig07} and \ref{fig08} show, respectively, the dependences of the magnetic field strength on the event horizon on the spin and the coefficient of radiation efficiency of SMBH. In these cases an explicit relationship between the parameters is not observed.

\subsection{Determination of magnetic field strength using the data obtained from the analysis of X-ray spectra} 

\begin{table*}
\label{tab02}
\footnotesize
\caption{Results of magnetic field strength assessment on the event horizon for objects for which the spin value is determined by the analysis of X-ray spectra.}
\setlength{\tabcolsep}{4pt}
\centering
\begin{tabular}{|l|c|c|l|r|r|r|r|}
\hline
Object & $\log\left(\frac{M_{BH}}{M_\odot}\right)$ & $a$ & FWHM  & $l_{E}$         & $\varepsilon  $ & $\log(k)$ & $\log(B_H)$ \\
\hline
MCG-6-30-15 & $6.65 \pm 0.17$ & $\geq 0.98$ & $1990 \pm 200$ & $0.40 \pm 0.13$ & $\geq 0.23$ & $-1.07^{+0.36}_{-0.32}$ & $4.74^{+0.26}_{-0.25}$ \\
Fairall 9 & $8.36 \pm 0.04$ & $0.52^{+0.19}_{-0.15}$ & $5618 \pm 107$ & $0.02 \pm 0.003$ & $0.08^{+0.02}_{-0.01}$ & $0.16^{+0.53}_{-0.51}$ & $3.89^{+0.29}_{-0.26}$ \\
SWIFT J2127.4+5654 & $7.18 \pm 0.07$ & $0.6^{+0.2}_{-0.2}$ & $2000 \pm 100$ & $0.18 \pm 0.03$ & $0.09^{+0.03}_{-0.02}$ & $-0.49^{+0.69}_{-0.59}$ & $4.92^{+0.36}_{-0.32}$ \\
1 H0707-495 & $6.60 \pm 0.06$ & $\geq 0.98$ & $940 \pm 47$ & $0.59^{+0.09}_{0.08}$ & $\geq 0.23$ & $-1.17^{+0.30}_{-0.29}$ & $5.77^{+0.20}_{-0.17}$ \\
Mrk 79 & $8.10 \pm 0.03$ & $0.7 \pm 0.1$ & $4735 \pm 44$ & $0.03 \pm 0.003$ & $0.10^{+0.02}_{-0.03}$ & $-0.18^{+0.35}_{-0.42}$ & $3.87^{+0.15}_{-0.20}$ \\
PG 0003+199 & $7.42 \pm 0.04$ & $0.70^{+0.12}_{-0.01}$ & $2182 \pm 53$ & $0.36^{+0.06}_{-0.05}$ & $0.10^{+0.02}_{-0.001}$ & $-0.79^{+0.26}_{-0.38}$ & $4.83^{+0.14}_{-0.20}$ \\
NGC 3783 & $7.60 \pm 0.03$ & $\geq 0.98$ & $3634 \pm 41$ & $0.06 \pm 0.006$ & $\geq 0.23$ & $-0.57^{+0.25}_{-0.24}$ & $4.15^{+0.14}_{-0.08}$ \\
Ark 120 & $8.67 \pm 0.03$ & $0.94 \pm 0.1$ & $5987 \pm 54$ & $0.02 \pm 0.002$ & $0.18^{+0.14}_{-0.05}$ & $-0.32^{+0.57}_{-0.49}$ & $3.74^{+0.26}_{-0.19}$ \\
3C 120 & $7.67 \pm 0.03$ & $\geq 0.95$ & $2419 \pm 29$ & $0.31^{+0.04}_{-0.03}$ & $\geq 0.19$ & $-0.99^{+0.37}_{-0.37}$ & $4.43^{+0.18}_{-0.11}$ \\
PG 0921+525 & $7.17^{+0.04}_{-0.05}$ & $\geq 0.99$ & $2194 \pm 64$ & $0.27^{+0.05}_{-0.04}$ & $\geq 0.26$ & $-0.97^{+0.24}_{-0.23}$ & $4.70^{+0.23}_{-0.16}$ \\
RBS 1124 & $8.26$ & $\geq 0.98$ & $4260 \pm 1250$ & $0.15$ & $\geq 0.23$ & $-0.82^{+0.25}_{-0.23}$ & $4.11^{+0.65}_{-0.52}$ \\
Mrk 359 & $6.68 \pm 0.06$ & $0.66^{+0.30}_{-0.54}$ & $900 \pm 45$ & $0.86^{+0.14}_{-0.11}$ & $0.10^{+0.10}_{-0.04}$ & $-0.97^{+2.07}_{-0.98}$ & $5.73^{+1.06}_{-0.45}$ \\
PG 1501+106 & $8.52 \pm 0.06$ & $\geq 0.52$ & $5300 \pm 265$ & $0.03^{+0.008}_{-0.006}$ & $\geq 0.08$ & $0.08^{+0.71}_{-1.19}$ & $3.87^{+0.41}_{-0.44}$ \\
Mrk 1018 & $8.15$ & $0.58^{+0.36}_{-0.74}$ & $6940 \pm 760$ & $0.01$ & $0.09^{+0.09}_{-0.04}$ & $0.27^{+1.56}_{-0.97}$ & $3.66^{+1.00}_{-0.56}$ \\
NGC 4051 & $6.05^{+0.06}_{-0.07}$ & $\geq 0.99$ & $1565 \pm 80$ & $0.32^{+0.08}_{-0.06}$ & $\geq 0.26$ & $-1.01^{+0.26}_{-0.25}$ & $5.08^{+0.18}_{-0.17}$ \\
NGC 1365 & $6.60^{+1.40}_{-0.30}$ & $0.97^{+0.01}_{-0.04}$ & $3586 \pm 179$ & $0.06^{+0.06}_{-0.04}$ & $0.21^{+0.02}_{-0.04}$ & $-0.58^{+0.51}_{-0.41}$ & $4.40^{+0.11}_{-0.30}$ \\
\hline
\end{tabular}
\end{table*}

In Table 2, we present the results of determination of the magnetic field strength on the event horizon for 16 AGNs for which the spin value of the central SMBH was obtained from their X-ray spectra \citep{brenneman13, reynolds14}. The calculations are performed using the relations (\ref{eq06}) and (\ref{eq10}). Inclination angle $i$ and $FWHM$ are taken from the review \citet{marin16}.

According to Table 2, the magnetic field strength on the event horizon ranges from 4.6 kG to 586 kG with a median value of about 26 kG. A histogram of the distribution of the number of SMBHs depending on the magnitude of the magnetic field on the event horizon calculated in the framework of the Meier hybrid model is shown in Fig.\ref{fig04}. The considered objects are shown by dense hatching.

In Figs.\ref{fig05}-\ref{fig08} the considered objects are shown by white circles.

Fig.\ref{fig05} illustrates the dependence of the magnetic field strength on the event horizon on the mass of the central SMBH. Despite a rather large scatter of points and uncertainties, an inverse relationship is traced $\log(B_H) \approx - (0.67 \pm 0.12) \log(M_{BH} / M_{\odot}) + (9.54 \pm 0.92)$. Fig.\ref{fig06} shows the dependence of the magnetic field strength on the event horizon on the Eddington ratio. A dependence of the form $\log(B_H) \approx (1.00 \pm 0.13) \log(l_E) + (5.44 \pm 0.14)$ is present. Figures \ref{fig07} and \ref{fig08} do not reveal any noticeable relationship between the magnetic field strength $B_{H}$ on the one hand and the spin $a$ and the radiation efficiency $\varepsilon$ on the other hand, respectively.

\subsection{Comparison of the results of calculations of the magnetic field strength in two samples} 

Fig.\ref{fig05} demonstrates the inverse relationship between the magnetic field strength on the event horizon $B_{H}$ and the mass of the central SMBH $M_{BH}$. We compared the samples considered in Section 3.2 (hereinafter referred to as sample 1) and Section 3.3 (hereinafter referred to as sample 2) for differences in their $M_ {BH}$ and $B_{H}$. Table 3 summarizes the basic statistical properties of these parameters for samples 1 and 2. Table 3 shows that sample 1 has, on average, a larger mass-volume number of data than sample 2; for the magnetic field strength the opposite is true.

\begin{table}
\label{tab03}
\caption{Basic statistical properties of the samples. ''mean'' indicates the arithmetic mean, ''median'' the median value, ''sd'' the standard deviation.}
\centering
\begin{tabular}{|c|c|c|c|c|c|c|}
\hline
Sample & \multicolumn{3}{c|}{$\log\left(\frac{M_{BH}}{M_\odot}\right)$} & \multicolumn{3}{c|}{$\log(B_H)$} \\
\cline{2-7}
 & mean & median & sd & mean & median & sd \\
 \hline
1 & $8.32$ & $8.38$ & $0.77$ & $4.11$ & $4.05$ & $0.52$ \\
2 & $7.48$ & $7.51$ & $0.81$ & $4.49$ & $4.42$ & $0.66$ \\
\hline
\end{tabular}
\end{table}

Next, we examined the differences between the samples for their statistical significance. A preliminary analysis of the normality of the distribution of the parameters under study using the Shapiro-Wilk test showed that the masses of SMBHs in a logarithmic scale in samples 1 and 2 are distributed normally. The distribution of $\log(B_{H})$ in both samples does not deviate from normal. The Bartlett test showed that samples 1 and 2 also satisfy the condition of homogeneity of variances of the studied parameters. Thus, the conditions for the applicability of the t-criterion are satisfied. The t-test confirmed the statistical difference between samples 1 and 2 at a significance level of $0.05$ for both parameters: $\log \left (\frac{M_{BH}}{M_\odot}\right)$ and $\log(B_H)$.

Note that three objects (PG~0003+199, PG~0921+525 and PG~1501+106) are included in both sample 1 and sample 2. The results of determination of the magnetic field strength on the horizon $B_{H}$ given in Table 1 and Table 2 for these objects differ by less than 1.4 times and are within the error range. The maximum difference for the object PG~0003+199 reaches 1.4 times, while for PG~0921+525 it is 1.1 times, and for PG~1501+106 the results are almost identical. We emphasize that in the calculations for these objects we took the values of the spin of the central SMBH and the inclination angle obtained by different methods. Of course, the question remains, whether the good agreement of the results is a coincidence and further research with a larger number of objects is needed. However, this agreement itself increases the significance of the results of determination of the magnetic field strength on the event horizon for these three objects.

\section{Conclusions} 

Estimates based on observational data in the visible and X-ray wavelengths and on the Shakura-Sunyaev model of the disk showed that the magnetic field strength on the event horizon of most SMBHs in the AGNs range from several to tens of kG, with an approximate average value of $10^4$ G. At the same time, for individual objects the magnetic fields are either substantially smaller - of the order of 1 kG and less, or appreciably higher - of the order of 1 MG.

For three objects (PG~0003+199, PG~0921+525 and PG~1501+106), the magnetic field strength on the event horizon was determined by two methods based on both optical and X-ray observations. The obtained results are well consistent within error limits. For six objects (PG~0003+199, PG~0026+129, PG~0804+761, PG~0844+349, PG~0953+414 and PG~1613+658) the magnetic field strength is consistent within the error with that determined in \citet{daly19}. For the remaining objects, the statistical properties of the results are consistent with the literature data for objects of this type \citep{pariev03, baczko16, mikhailov15, daly19}.

When analyzing the data, we obtained the dependence of the magnetic field strength on the mass of SMBH in the form $\log(B_H) \approx - (0.68 \pm 0.04) \log(M_{BH} / M_{\odot}) + (9.69 \pm 0.32) $ and $\log(B_H) \approx - (0.67 \pm 0.12) \log(M_{BH} / M_{\odot}) + (9.54 \pm 0.92)$ for sample 1 and sample 2, respectively, which is in good agreement with the mass dependence of the Eddington magnetic field of the accretion disk: $\log(B_{Edd}) \approx - 0.5 \log(M_{BH} / M_{\odot}) + 8.77$ \citep{daly19}. We also derived the dependence of the magnetic field strength on the event horizon on the Eddington ratio in the form $\log(B_H) \approx (0.65 \pm 0.04) \log(l_E) + (4.92 \pm 0.06) $ and $\log(B_H) \approx (1.00 \pm 0.13) \log(l_E) + (5.44 \pm 0.14)$ for sample 1 and sample 2, respectively.

In general, the problem of determination of magnetic field strength on the SMBH event horizon requires further research. In particular, a method of direct determination of the magnetic field strength at the radius of the innermost stable circular orbit (i.e., the inner radius of the accretion disk) from the characteristics of the radiation from this and adjacent regions of the disk should be developed, based on spectroscopic and, if possible, polarimetric observations in the ultraviolet and X-ray spectral ranges. For example, one should take into account such an effect as Faraday rotation of polarization plane in strong magnetic field, which can lead to a characteristic form of the dependence of the degree of polarization on the wavelength, or even depolarization of radiation from inner region of the accretion disk. Such observations may become possible with the launch of the space observatories ''Spektr-UV'', ''Spektr-RG'' and IXPE.

\section*{Acknowledgements}

This research was partially supported by RAS program of basic research. S.D. Buliga was supported by the Russian Science Foundation (project No. 20-12-00030 ''Investigation of geometry and kinematics of ionized gas in active galactic nuclei by polarimetry methods'').

\bibliographystyle{mnras}
\bibliography{mybibfile}

\begin{thebibliography}{}
\makeatletter
\relax
\def\mn@urlcharsother{\let\do\@makeother \do\$\do\&\do\#\do\^\do\_\do\%\do\~}
\def\mn@doi{\begingroup\mn@urlcharsother \@ifnextchar [ {\mn@doi@}
  {\mn@doi@[]}}
\def\mn@doi@[#1]#2{\def\@tempa{#1}\ifx\@tempa\@empty \href
  {http://dx.doi.org/#2} {doi:#2}\else \href {http://dx.doi.org/#2} {#1}\fi
  \endgroup}
\def\mn@eprint#1#2{\mn@eprint@#1:#2::\@nil}
\def\mn@eprint@arXiv#1{\href {http://arxiv.org/abs/#1} {{\tt arXiv:#1}}}
\def\mn@eprint@dblp#1{\href {http://dblp.uni-trier.de/rec/bibtex/#1.xml}
  {dblp:#1}}
\def\mn@eprint@#1:#2:#3:#4\@nil{\def\@tempa {#1}\def\@tempb {#2}\def\@tempc
  {#3}\ifx \@tempc \@empty \let \@tempc \@tempb \let \@tempb \@tempa \fi \ifx
  \@tempb \@empty \def\@tempb {arXiv}\fi \@ifundefined
  {mn@eprint@\@tempb}{\@tempb:\@tempc}{\expandafter \expandafter \csname
  mn@eprint@\@tempb\endcsname \expandafter{\@tempc}}}

\bibitem[\protect\citeauthoryear{{Afanasiev}, {Gnedin}, {Piotrovich},
  {Natsvlishvili}  \& {Buliga}}{{Afanasiev} et~al.}{2018}]{afanasiev18}
{Afanasiev} V.~L.,  {Gnedin} Y.~N.,  {Piotrovich} M.~Y.,  {Natsvlishvili}
  T.~M.,   {Buliga} S.~D.,  2018, \mn@doi [Astronomy Letters]
  {10.1134/S1063773718060014}, 44, 362

\bibitem[\protect\citeauthoryear{{Baczko} et~al.,}{{Baczko}
  et~al.}{2016}]{baczko16}
{Baczko} A.-K.,  et~al., 2016, \mn@doi [\aap] {10.1051/0004-6361/201527951},
  \href {https://ui.adsabs.harvard.edu/abs/2016A%26A...593A..47B} {593, A47}

\bibitem[\protect\citeauthoryear{{Bardeen}, {Press}  \& {Teukolsky}}{{Bardeen}
  et~al.}{1972}]{bardeen72}
{Bardeen} J.~M.,  {Press} W.~H.,   {Teukolsky} S.~A.,  1972, \mn@doi [\apj]
  {10.1086/151796}, \href {http://adsabs.harvard.edu/abs/1972ApJ...178..347B}
  {178, 347}

\bibitem[\protect\citeauthoryear{{Bentz} et~al.,}{{Bentz}
  et~al.}{2013}]{bentz13}
{Bentz} M.~C.,  et~al., 2013, \mn@doi [\apj] {10.1088/0004-637X/767/2/149},
  \href {http://adsabs.harvard.edu/abs/2013ApJ...767..149B} {767, 149}

\bibitem[\protect\citeauthoryear{{Blandford} \& {Payne}}{{Blandford} \&
  {Payne}}{1982}]{blandford82}
{Blandford} R.~D.,  {Payne} D.~G.,  1982, \mn@doi [\mnras]
  {10.1093/mnras/199.4.883}, \href
  {http://adsabs.harvard.edu/abs/1982MNRAS.199..883B} {199, 883}

\bibitem[\protect\citeauthoryear{{Blandford} \& {Znajek}}{{Blandford} \&
  {Znajek}}{1977}]{blandford77}
{Blandford} R.~D.,  {Znajek} R.~L.,  1977, \mn@doi [\mnras]
  {10.1093/mnras/179.3.433}, \href
  {http://adsabs.harvard.edu/abs/1977MNRAS.179..433B} {179, 433}

\bibitem[\protect\citeauthoryear{{Brenneman}}{{Brenneman}}{2013}]{brenneman13}
{Brenneman} L.,  2013, Acta Polytechnica, \href
  {http://adsabs.harvard.edu/abs/2013AcPol..53..652B} {53, 652}

\bibitem[\protect\citeauthoryear{{Capellupo}, {Netzer}, {Lira}, {Trakhtenbrot}
  \& {Mej{\'\i}a-Restrepo}}{{Capellupo} et~al.}{2016}]{capellupo16}
{Capellupo} D.~M.,  {Netzer} H.,  {Lira} P.,  {Trakhtenbrot} B.,
  {Mej{\'\i}a-Restrepo} J.,  2016, \mn@doi [\mnras] {10.1093/mnras/stw937},
  \href {https://ui.adsabs.harvard.edu/abs/2016MNRAS.460..212C} {460, 212}

\bibitem[\protect\citeauthoryear{{Chandrasekhar}}{{Chandrasekhar}}{1950}]{chandrasekhar50}
{Chandrasekhar} S.,  1950, {Radiative transfer.}.
Clarendon Press, Oxford

\bibitem[\protect\citeauthoryear{{Condon}, {Cotton}, {Greisen}, {Yin},
  {Perley}, {Taylor}  \& {Broderick}}{{Condon} et~al.}{1998}]{condon98}
{Condon} J.~J.,  {Cotton} W.~D.,  {Greisen} E.~W.,  {Yin} Q.~F.,  {Perley}
  R.~A.,  {Taylor} G.~B.,   {Broderick} J.~J.,  1998, \mn@doi [\aj]
  {10.1086/300337}, \href
  {https://ui.adsabs.harvard.edu/abs/1998AJ....115.1693C} {115, 1693}

\bibitem[\protect\citeauthoryear{{Daly}}{{Daly}}{2019}]{daly19}
{Daly} R.~A.,  2019, preprint, \href
  {https://ui.adsabs.harvard.edu/abs/2019arXiv190511319D} {} (\mn@eprint
  {arXiv} {1905.11319})

\bibitem[\protect\citeauthoryear{{Daly}, {Stout}  \& {Mysliwiec}}{{Daly}
  et~al.}{2018}]{daly16}
{Daly} R.~A.,  {Stout} D.~A.,   {Mysliwiec} J.~N.,  2018, \mn@doi [\apj]
  {10.3847/1538-4357/aad08b}, \href
  {https://ui.adsabs.harvard.edu/abs/2018ApJ...863..117D} {863, 117}

\bibitem[\protect\citeauthoryear{{Decarli}, {Dotti}  \& {Treves}}{{Decarli}
  et~al.}{2011}]{decarli11}
{Decarli} R.,  {Dotti} M.,   {Treves} A.,  2011, \mn@doi [\mnras]
  {10.1111/j.1365-2966.2010.18102.x}, \href
  {http://adsabs.harvard.edu/abs/2011MNRAS.413...39D} {413, 39}

\bibitem[\protect\citeauthoryear{{Du} et~al.,}{{Du} et~al.}{2014}]{du14}
{Du} P.,  et~al., 2014, \mn@doi [\apj] {10.1088/0004-637X/782/1/45}, \href
  {http://adsabs.harvard.edu/abs/2014ApJ...782...45D} {782, 45}

\bibitem[\protect\citeauthoryear{{Edelson} et~al.,}{{Edelson}
  et~al.}{2015}]{edelson15}
{Edelson} R.,  et~al., 2015, \mn@doi [\apj] {10.1088/0004-637X/806/1/129},
  \href {https://ui.adsabs.harvard.edu/abs/2015ApJ...806..129E} {806, 129}

\bibitem[\protect\citeauthoryear{{Feng}, {Shen}  \& {Li}}{{Feng}
  et~al.}{2014}]{feng14}
{Feng} H.,  {Shen} Y.,   {Li} H.,  2014, \mn@doi [\apj]
  {10.1088/0004-637X/794/1/77}, \href
  {http://adsabs.harvard.edu/abs/2014ApJ...794...77F} {794, 77}

\bibitem[\protect\citeauthoryear{{Foschini}}{{Foschini}}{2011}]{foschini11}
{Foschini} L.,  2011, \mn@doi [Research in Astronomy and Astrophysics]
  {10.1088/1674-4527/11/11/003}, \href
  {https://ui.adsabs.harvard.edu/abs/2011RAA....11.1266F} {11, 1266}

\bibitem[\protect\citeauthoryear{{Gnedin}}{{Gnedin}}{2013}]{gnedin13b}
{Gnedin} Y.~N.,  2013, \mn@doi [Physics Uspekhi]
  {10.3367/UFNe.0183.201307f.0747}, \href
  {https://ui.adsabs.harvard.edu/abs/2013PhyU...56..709G} {56, 709}

\bibitem[\protect\citeauthoryear{{Gnedin}, {Buliga}, {Silant'ev},
  {Natsvlishvili}  \& {Piotrovich}}{{Gnedin} et~al.}{2012}]{gnedin12}
{Gnedin} Y.~N.,  {Buliga} S.~D.,  {Silant'ev} N.~A.,  {Natsvlishvili} T.~M.,
  {Piotrovich} M.~Y.,  2012, \mn@doi [\apss] {10.1007/s10509-012-1146-y}, \href
  {https://ui.adsabs.harvard.edu/abs/2012Ap%26SS.342..137G} {342, 137}

\bibitem[\protect\citeauthoryear{{Gnedin}, {Piotrovich}, {Silant'ev},
  {Natsvlishvili}  \& {Buliga}}{{Gnedin} et~al.}{2015}]{gnedin15}
{Gnedin} Y.~N.,  {Piotrovich} M.~Y.,  {Silant'ev} N.~A.,  {Natsvlishvili}
  T.~M.,   {Buliga} S.~D.,  2015, \mn@doi [Astrophysics]
  {10.1007/s10511-015-9398-1}, \href
  {http://adsabs.harvard.edu/abs/2015Ap.....58..443G} {58, 443}

\bibitem[\protect\citeauthoryear{{Grier}, {Pancoast}, {Barth}, {Fausnaugh},
  {Brewer}, {Treu}  \& {Peterson}}{{Grier} et~al.}{2017}]{grier17}
{Grier} C.~J.,  {Pancoast} A.,  {Barth} A.~J.,  {Fausnaugh} M.~M.,  {Brewer}
  B.~J.,  {Treu} T.,   {Peterson} B.~M.,  2017, \mn@doi [\apj]
  {10.3847/1538-4357/aa901b}, \href
  {http://adsabs.harvard.edu/abs/2017ApJ...849..146G} {849, 146}

\bibitem[\protect\citeauthoryear{{Grupe} \& {Nousek}}{{Grupe} \&
  {Nousek}}{2015}]{grupe15}
{Grupe} D.,  {Nousek} J.~A.,  2015, \mn@doi [\aj] {10.1088/0004-6256/149/2/85},
  \href {https://ui.adsabs.harvard.edu/abs/2015AJ....149...85G} {149, 85}

\bibitem[\protect\citeauthoryear{{Homayouni} et~al.,}{{Homayouni}
  et~al.}{2019}]{homayouni19}
{Homayouni} Y.,  et~al., 2019, \mn@doi [\apj] {10.3847/1538-4357/ab2638}, \href
  {https://ui.adsabs.harvard.edu/abs/2019ApJ...880..126H} {880, 126}

\bibitem[\protect\citeauthoryear{{Kokubo}}{{Kokubo}}{2018}]{kokubo18}
{Kokubo} M.,  2018, \mn@doi [\pasj] {10.1093/pasj/psy096}, \href
  {https://ui.adsabs.harvard.edu/abs/2018PASJ...70...97K} {70, 97}

\bibitem[\protect\citeauthoryear{{Kollatschny}, {Ochmann}, {Zetzl}, {Haas},
  {Chelouche}, {Kaspi}, {Pozo Nu{\~n}ez}  \& {Grupe}}{{Kollatschny}
  et~al.}{2018}]{kollatschny18}
{Kollatschny} W.,  {Ochmann} M.~W.,  {Zetzl} M.,  {Haas} M.,  {Chelouche} D.,
  {Kaspi} S.,  {Pozo Nu{\~n}ez} F.,   {Grupe} D.,  2018, \mn@doi [\aap]
  {10.1051/0004-6361/201833727}, \href
  {https://ui.adsabs.harvard.edu/abs/2018A%26A...619A.168K} {619, A168}

\bibitem[\protect\citeauthoryear{{Laor} \& {Davis}}{{Laor} \&
  {Davis}}{2014}]{laor14}
{Laor} A.,  {Davis} S.~W.,  2014, \mn@doi [\mnras] {10.1093/mnras/stt2408},
  \href {https://ui.adsabs.harvard.edu/abs/2014MNRAS.438.3024L} {438, 3024}

\bibitem[\protect\citeauthoryear{{Li}}{{Li}}{2002}]{li02}
{Li} L.-X.,  2002, \mn@doi [\apj] {10.1086/338486}, \href
  {http://adsabs.harvard.edu/abs/2002ApJ...567..463L} {567, 463}

\bibitem[\protect\citeauthoryear{{Ma}, {Yuan}  \& {Wang}}{{Ma}
  et~al.}{2007}]{ma07}
{Ma} R.-Y.,  {Yuan} F.,   {Wang} D.-X.,  2007, \mn@doi [\apj] {10.1086/522917},
  \href {http://adsabs.harvard.edu/abs/2007ApJ...671.1981M} {671, 1981}

\bibitem[\protect\citeauthoryear{{Marin}}{{Marin}}{2016}]{marin16}
{Marin} F.,  2016, \mn@doi [\mnras] {10.1093/mnras/stw1131}, \href
  {http://adsabs.harvard.edu/abs/2016MNRAS.460.3679M} {460, 3679}

\bibitem[\protect\citeauthoryear{{Meier}}{{Meier}}{1999}]{meier99}
{Meier} D.~L.,  1999, \mn@doi [\apj] {10.1086/307671}, \href
  {http://adsabs.harvard.edu/abs/1999ApJ...522..753M} {522, 753}

\bibitem[\protect\citeauthoryear{{Merloni} \& {Heinz}}{{Merloni} \&
  {Heinz}}{2007}]{merloni07}
{Merloni} A.,  {Heinz} S.,  2007, \mn@doi [\mnras]
  {10.1111/j.1365-2966.2007.12253.x}, \href
  {http://adsabs.harvard.edu/abs/2007MNRAS.381..589M} {381, 589}

\bibitem[\protect\citeauthoryear{{Mikhailov}, {Gnedin}  \&
  {Belonovsky}}{{Mikhailov} et~al.}{2015}]{mikhailov15}
{Mikhailov} A.~G.,  {Gnedin} Y.~N.,   {Belonovsky} A.~V.,  2015, \mn@doi
  [Astrophysics] {10.1007/s10511-015-9372-y}, \href
  {https://ui.adsabs.harvard.edu/abs/2015Ap.....58..157M} {58, 157}

\bibitem[\protect\citeauthoryear{{Moderski}, {Sikora}  \& {Lasota}}{{Moderski}
  et~al.}{1997}]{moderski97}
{Moderski} R.,  {Sikora} M.,   {Lasota} J.~P.,  1997, in {Ostrowski} M.,
  {Sikora} M.,  {Madejski} G.,   {Begelman} M.,  eds, Relativistic Jets in
  AGNs. pp 110--116 (\mn@eprint {arXiv} {astro-ph/9706263})

\bibitem[\protect\citeauthoryear{{Pariev}, {Blackman}  \& {Boldyrev}}{{Pariev}
  et~al.}{2003}]{pariev03}
{Pariev} V.~I.,  {Blackman} E.~G.,   {Boldyrev} S.~A.,  2003, \mn@doi [\aap]
  {10.1051/0004-6361:20030868}, \href
  {https://ui.adsabs.harvard.edu/abs/2003A&A...407..403P} {407, 403}

\bibitem[\protect\citeauthoryear{{Reynolds}}{{Reynolds}}{2014}]{reynolds14}
{Reynolds} C.~S.,  2014, \mn@doi [\ssr] {10.1007/s11214-013-0006-6}, \href
  {http://adsabs.harvard.edu/abs/2014SSRv..183..277R} {183, 277}

\bibitem[\protect\citeauthoryear{{Sani}, {Lutz}, {Risaliti}, {Netzer}, {Gallo},
  {Trakhtenbrot}, {Sturm}  \& {Boller}}{{Sani} et~al.}{2010}]{sani10}
{Sani} E.,  {Lutz} D.,  {Risaliti} G.,  {Netzer} H.,  {Gallo} L.~C.,
  {Trakhtenbrot} B.,  {Sturm} E.,   {Boller} T.,  2010, \mn@doi [\mnras]
  {10.1111/j.1365-2966.2009.16217.x}, \href
  {https://ui.adsabs.harvard.edu/abs/2010MNRAS.403.1246S} {403, 1246}

\bibitem[\protect\citeauthoryear{{Shakura} \& {Sunyaev}}{{Shakura} \&
  {Sunyaev}}{1973}]{shakura73}
{Shakura} N.~I.,  {Sunyaev} R.~A.,  1973, \aap, \href
  {http://adsabs.harvard.edu/abs/1973A%26A....24..337S} {24, 337}

\bibitem[\protect\citeauthoryear{{Sheinis} \&
  {L{\'o}pez-S{\'a}nchez}}{{Sheinis} \&
  {L{\'o}pez-S{\'a}nchez}}{2017}]{sheinis17}
{Sheinis} A.~I.,  {L{\'o}pez-S{\'a}nchez} {\'A}.~R.,  2017, \mn@doi [\aj]
  {10.3847/1538-3881/153/2/55}, \href
  {https://ui.adsabs.harvard.edu/abs/2017AJ....153...55S} {153, 55}

\bibitem[\protect\citeauthoryear{{Slone} \& {Netzer}}{{Slone} \&
  {Netzer}}{2012}]{slone12}
{Slone} O.,  {Netzer} H.,  2012, \mn@doi [\mnras]
  {10.1111/j.1365-2966.2012.21699.x}, \href
  {https://ui.adsabs.harvard.edu/abs/2012MNRAS.426..656S} {426, 656}

\bibitem[\protect\citeauthoryear{{Sobolev}}{{Sobolev}}{1963}]{sobolev63}
{Sobolev} V.~V.,  1963, {A treatise on radiative transfer.}.
Van Nostrand, Princeton, N.J.

\bibitem[\protect\citeauthoryear{{Sun}, {Xue}, {Trump}  \& {Gu}}{{Sun}
  et~al.}{2019}]{sun19}
{Sun} M.,  {Xue} Y.,  {Trump} J.~R.,   {Gu} W.-M.,  2019, \mn@doi [\mnras]
  {10.1093/mnras/sty2885}, \href
  {https://ui.adsabs.harvard.edu/abs/2019MNRAS.482.2788S} {482, 2788}

\bibitem[\protect\citeauthoryear{{Trakhtenbrot}, {Volonteri}  \&
  {Natarajan}}{{Trakhtenbrot} et~al.}{2017}]{trakhtenbrot17}
{Trakhtenbrot} B.,  {Volonteri} M.,   {Natarajan} P.,  2017, \mn@doi [\apjl]
  {10.3847/2041-8213/836/1/L1}, \href
  {https://ui.adsabs.harvard.edu/abs/2017ApJ...836L...1T} {836, L1}

\bibitem[\protect\citeauthoryear{{Vestergaard} \& {Peterson}}{{Vestergaard} \&
  {Peterson}}{2006}]{vestergaard06}
{Vestergaard} M.,  {Peterson} B.~M.,  2006, \mn@doi [\apj] {10.1086/500572},
  \href {http://adsabs.harvard.edu/abs/2006ApJ...641..689V} {641, 689}

\bibitem[\protect\citeauthoryear{{Wang}, {Xiao}  \& {Lei}}{{Wang}
  et~al.}{2002}]{wang02}
{Wang} D.~X.,  {Xiao} K.,   {Lei} W.~H.,  2002, \mn@doi [\mnras]
  {10.1046/j.1365-8711.2002.05652.x}, \href
  {http://adsabs.harvard.edu/abs/2002MNRAS.335..655W} {335, 655}

\bibitem[\protect\citeauthoryear{{Wang}, {Ma}, {Lei}  \& {Yao}}{{Wang}
  et~al.}{2003}]{wang03}
{Wang} D.-X.,  {Ma} R.-Y.,  {Lei} W.-H.,   {Yao} G.-Z.,  2003, \mn@doi [\apj]
  {10.1086/377303}, \href {http://adsabs.harvard.edu/abs/2003ApJ...595..109W}
  {595, 109}

\bibitem[\protect\citeauthoryear{{Yu} et~al.,}{{Yu} et~al.}{2020}]{yu20}
{Yu} Z.,  et~al., 2020, \mn@doi [\apjs] {10.3847/1538-4365/ab5e7a}, \href
  {https://ui.adsabs.harvard.edu/abs/2020ApJS..246...16Y} {246, 16}

\bibitem[\protect\citeauthoryear{{Zhang}, {Lu}  \& {Zhang}}{{Zhang}
  et~al.}{2005}]{zhang05}
{Zhang} W.-M.,  {Lu} Y.,   {Zhang} S.-N.,  2005, Chinese Journal of Astronomy
  and Astrophysics Supplement, \href
  {http://adsabs.harvard.edu/abs/2005ChJAS...5..347Z} {5, 347}

\makeatother
\end{thebibliography}

\bsp
\label{lastpage}
\end{document}